\documentclass{article}
\usepackage{spconf,amsmath,graphicx,hyperref}
\usepackage{booktabs}
\usepackage{soul}
\usepackage{color}
\usepackage{booktabs}
\usepackage{threeparttable}
\usepackage{fancyhdr}

\title{Spike Encoding for Environmental Sound: A Comparative Benchmark}

\name{
\begin{minipage}{\textwidth}\centering
Andres Larroza\textsuperscript{1} \quad
Javier Naranjo-Alcazar\textsuperscript{1} \quad
Vicent Ortiz-Castello\textsuperscript{1} \quad
Maximo Cobos\textsuperscript{2} \quad
Pedro Zuccarello\textsuperscript{1}\thanks{The participation of all the researchers affiliated at ITI in this work was funded by the Valencian Institute for Business Competitiveness (IVACE) and the FEDER funds by means of project LIASound (IMDEEA/2024/110). The research carried out for this publication has been partially funded by the project STARRING-NEURO (PID2022-137048OA-C44) funded by the Spanish State Research Agency and the European Union MCIN/AEI/10.13039/501100011033/FEDER, UE.}
\end{minipage}
}

\address{
\textsuperscript{1}Audio and Neuromorphic Processing Group, Instituto Tecnologico de Informatica (ITI), Valencia, Spain \\
\textsuperscript{2}Department of Computer Science, Universitat de Valencia, 46100 Burjassot, Spain
}

\begin{document}

\maketitle

\begin{abstract}
Spiking Neural Networks (SNNs) offer energy‑efficient processing suitable for edge applications, but conventional sensor data must first be converted into spike trains for neuromorphic processing. Environmental sound—including urban soundscapes—poses challenges due to variable frequencies, background noise, and overlapping acoustic events, while most spike‑based audio encoding research has focused on speech. This paper analyzes three spike encoding methods—Threshold Adaptive Encoding (TAE), Step Forward (SF), and Moving Window (MW)—across three datasets: ESC‑10, UrbanSound8K, and TAU Urban Acoustic Scenes. Our multi‑band analysis shows that TAE consistently outperforms SF and MW in reconstruction quality, both per frequency band and per class across datasets. Moreover, TAE yields the lowest spike firing rates, indicating superior energy efficiency. For downstream environmental sound classification with a standard SNN, TAE also achieves the best performance among the compared encoders. Overall, this work provides foundational insights and a comparative benchmark to guide the selection of spike encoders for neuromorphic environmental sound processing.
\end{abstract}


\section{Introduction}
\label{sec:intro}

Spiking Neural Networks (SNNs) have emerged as a promising paradigm in neuromorphic computing due to their energy efficiency and biologically inspired processing mechanisms \cite{Malcolm2023}. Unlike traditional artificial neural networks (ANNs), which rely on continuous-valued activations \cite{Eshraghian2023}, SNNs process information through discrete spike events, mimicking the behavior of biological neurons \cite{Zhang2022}. To harness SNNs for real-world applications, sensor data need to be represented as spike trains. While neuromorphic sensors are specifically designed to generate spike-based outputs directly \cite{Lichsteiner:2008:JSSC}, conventional sensors typically require an additional encoding step. In this context, various spike encoding methods have been proposed, each with distinct characteristics influencing the efficiency and accuracy of SNN-based models \cite{Auge2021}. The selection of an appropriate encoding scheme is crucial, as it directly affects the network's capability to extract meaningful temporal patterns from input signals \cite{Petro2020}.


Although spike encoding techniques have been extensively studied in the context of speech processing \cite{Stewart2023}, their application to environmental sound processing remains underexplored. Environmental sounds, such as traffic noise, alarms, and natural ambient sounds, exhibit complex spectral-temporal structures, often containing overlapped acoustic events and substantial background noise \cite{Springer2024}. Unlike speech, which follows structured linguistic patterns, environmental sounds are highly unstructured and vary significantly in duration and frequency content. These unique challenges require a deeper investigation into the suitability of different spike encoding schemes for environmental sound processing.

State-of-the-art environmental sound classification methods predominantly employ convolutional neural networks (CNNs) or other machine learning models applied to spectrogram representations, which effectively capture the inherent time-frequency characteristics of audio signals \cite{guzhov2021esresnet, kong2020panns, singh2024atgnn, pellegrini2023adapting}. However, to the best of our knowledge, no state-of-the-art solution has yet encoded environmental sound datasets using spike-based methods for SNNs. This gap presents a promising opportunity to investigate neuromorphic-inspired approaches that may offer good temporal resolution and enhanced energy efficiency.

In this study, we conduct a comparative evaluation of widely used spike encoding methods to assess their performance in environmental sound processing. We analyze how these approaches preserve signal characteristics across different frequency ranges and sound categories, while quantifying their impact on energy efficiency and classification performance within SNNs. Our primary objective is to establish a comprehensive benchmark that offers foundational guidance for encoder selection, addresses the current gap in spike-based processing of non-speech audio, and promotes broader adoption of neuromorphic approaches in audio applications.

\section{Datasets and Methodology}
\label{sec:methods}

\subsection{Datasets}
\label{ssec:datasets}

We analyzed spike encoders across three complementary datasets that capture different aspects of environmental sounds, ranging from isolated events to complex acoustic scenes.

\textbf{ESC-10}: A subset of ESC-50 containing 400 recordings, divided into 10 classes with 40 clips per class. Each clip has a duration of 5~s. The dataset provides a balanced representation of environmental events, including dog barking, rain, sea waves, crying baby, clock ticking, sneezing, helicopter, chainsaw, crowing rooster, and fire crackling. We used all original 5~s recordings without modification \cite{Piczak2015}.

\textbf{UrbanSound8K}: The original dataset contains 8{,}732 recordings distributed across 10 classes (air conditioner, car horn, children playing, dog bark, drilling, engine idling, gun shot, jackhammer, siren, and street music), each clip lasting up to 4~s. To standardize audio length and ensure a fair comparison in the classification analysis, we retained only clips of exactly 4~s. Having the same lengths avoids padding or other manipulations that could bias the results. We also maintained class balance across folds. However, due to an insufficient number of valid 4~s samples, the classes \emph{drilling} and \emph{car horn} were removed, resulting in a filtered subset containing 8 classes and 3{,}768 clips \cite{salamon2014urbansound}.

\textbf{TAU-3Class}: Derived from the DCASE 2020 Acoustic Scene Classification task, we used recordings from device~A grouped into three categories: indoor, outdoor, and transportation. Each original recording is 10~s. To match lengths with the other datasets, we extracted the central 5~s segment, ensuring retention of the most relevant acoustic information. We subsampled the official development split to balance the classes and simplify the analysis, yielding 1{,}500 training and 300 test clips (1{,}800 total) \cite{Mesaros2018}.\\

The complementary nature of these datasets allows evaluation across multiple scales of acoustic complexity: ESC-10 and UrbanSound8K emphasize discrimination of individual sources, whereas TAU-3Class focuses on coarser scene-level classification.

For the methods presented in the following sections, signal reconstruction and multi-band analysis were performed on the official test split of TAU-3Class, and on the first fold of the cross-validation partitions for ESC-10 and UrbanSound8K.

\subsection{Signal Reconstruction and Multi-band Analysis}
\label{ssec:processing}

Audio signals were processed as 128-band Mel-spectrograms spanning 20–20,000 Hz, computed using a 1024-point FFT with a hop length of 256 samples. These spectrograms served as input features for the spike encoders, with each Mel band encoded using the selected spike encoding method. 

The Mel-spectrogram was chosen for its computational efficiency and demonstrated effectiveness \cite{Stewart2023}. While prior work has employed Gammatone or Butterworth filters to compute cochleagrams \cite{ElFerdaoussi2023, Forno2022}, our tests indicated higher computational costs for high-frequency content; the Mel-spectrogram provided a more efficient solution for environmental sound analysis.  

To evaluate reconstruction quality, the 128 Mel bins were grouped into eight contiguous frequency bands: 20–125 Hz, 125–250 Hz, 250–500 Hz, 500 Hz–1 kHz, 1–2 kHz, 2–4 kHz, 4–8 kHz, and 8–20 kHz. Each band aggregated all Mel bins with center frequencies within the specified range.  

Reconstruction quality was assessed using two metrics computed per frequency band—ERRdB (error in decibels) and SNR (signal-to-noise ratio). In addition, ERRdB was computed per sound class for finer-grained analysis. Signal reconstruction was performed using the decoding methods described in \cite{Petro2020}.

\subsection{Spike Encoding Methods}
\label{ssec:encoding}

Spiking neural networks (SNNs) require transforming continuous signals into discrete spike trains, a process known as spike encoding. The choice of encoding method strongly influences downstream performance. In this study, three techniques—Moving Window (MW), Step Forward (SF), and Threshold Adaptive Encoding (TAE)—were selected based on demonstrated effectiveness in prior work \cite{Forno2022, Petro2020, Wang2024}, yet remain underexplored for environmental sound classification, motivating their evaluation here.

Spike encoders were applied channel-wise to the outputs of the Mel-spectrogram representation. All algorithms were implemented in Python (v3.10.12), following the pseudocode provided in the referenced works.

\textbf{Moving Window (MW):} Partitions the signal into overlapping windows and emits spikes based on intra-window signal variation, capturing local temporal patterns \cite{Forno2022}.

\textbf{Step Forward (SF):} Generates spikes when the difference between consecutive samples exceeds a predefined threshold, preserving salient changes while suppressing minor fluctuations \cite{Forno2022}.

\textbf{Threshold Adaptive Encoding (TAE):} Adapts the spike-generation threshold to signal characteristics, improving robustness to amplitude variability and sensitivity to significant changes \cite{Wang2024}.

\subsection{SNN Classification}
\label{ssec:snn}

To evaluate classification accuracy, we implemented a spiking neural network (SNN) with four fully connected layers and Leaky Integrate-and-Fire (LIF) neurons using snntorch (v0.9.1) \cite{Eshraghian2023}. The architecture comprises an input layer matching the Mel-spectrogram channels, three hidden layers with 128 neurons each, and an output layer matching the number of classes. Training was performed with a batch size of 32, a learning rate of 0.01, and 100 epochs.

Classification performance was reported using macro-averaged accuracy to provide a balanced assessment across classes. Baseline results from the original dataset papers are included for context. The goal is not to surpass these baselines, but to establish a comprehensive benchmark of spike encoding techniques for environmental sound processing; baselines are provided solely for reference.

Evaluations adhered to the original dataset protocols and partitions: 5-fold cross-validation for ESC-10, 10-fold cross-validation for UrbanSound8K, and the predefined train/test split for TAU-3Class.

\section{Results}
\label{sec:results}

\subsection{Signal Reconstruction}
\label{ssec:sigrecons}

\begin{figure}[tb]
\centering
\includegraphics[width=1.0\linewidth]{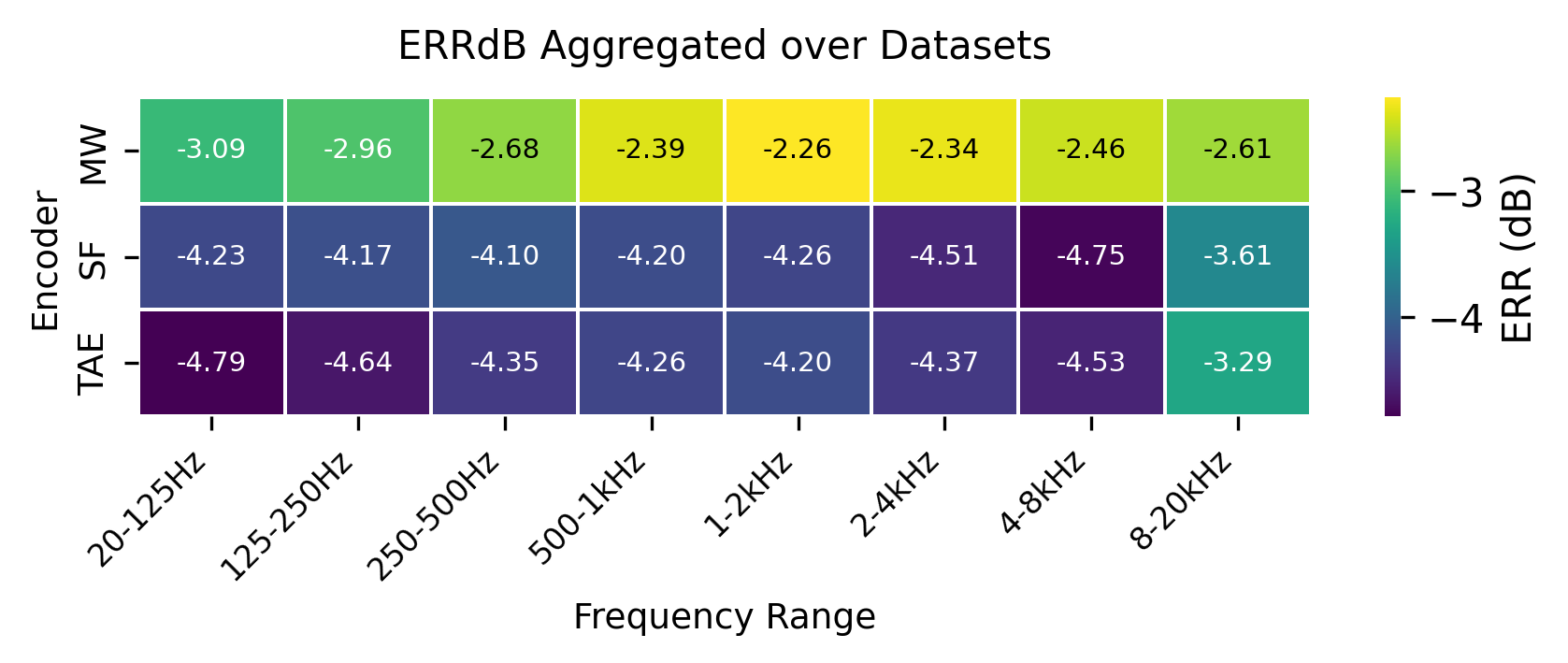}
\includegraphics[width=1.0\linewidth]{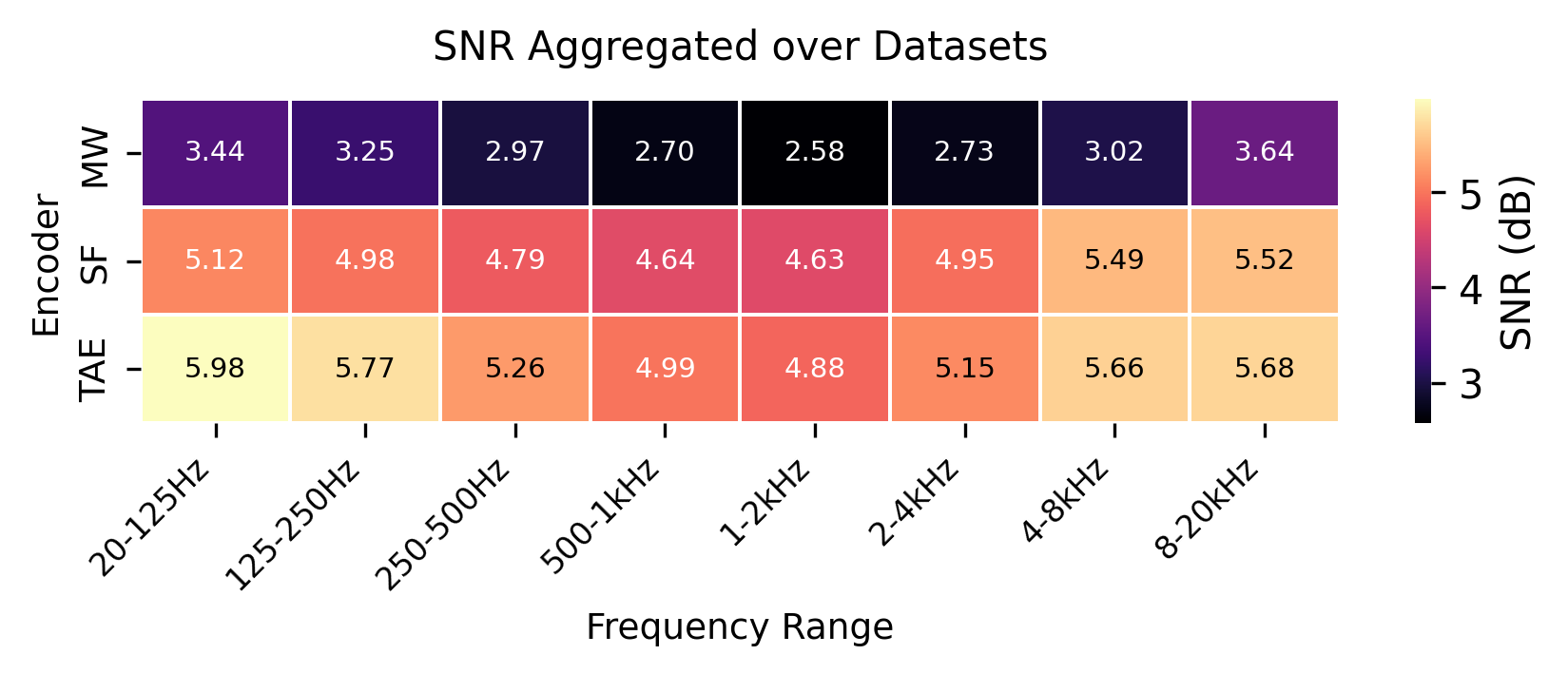}
\caption{Error in decibels (ERRdB) and signal-to-noise ratio (SNR) per frequency band, aggregated across all datasets. MW: Moving Window, SF: Step Forward, TAE: Threshold Adaptive Encoder.}
\label{fig:ERRdB-SNR-frequencies}
\end{figure}

Figure~\ref{fig:ERRdB-SNR-frequencies} summarizes multi-band reconstruction error and SNR performance aggregated over all datasets. TAE consistently yields the lowest reconstruction errors (most negative ERRdB) across nearly all frequency bands, particularly in the lower and mid ranges. Step Forward (SF) matches or slightly surpasses TAE only in the highest bands, while Moving Window (MW) shows the highest ERRdB values consistently. Notably, reconstruction error increases at higher frequencies for all encoders, emphasizing the difficulty of capturing rapid spectral changes. Corresponding SNR results mirror these trends: TAE achieves the highest SNR, SF is intermediate, and MW is lowest.

\begin{figure}[tb]
\centering
\includegraphics[width=1.0\linewidth]{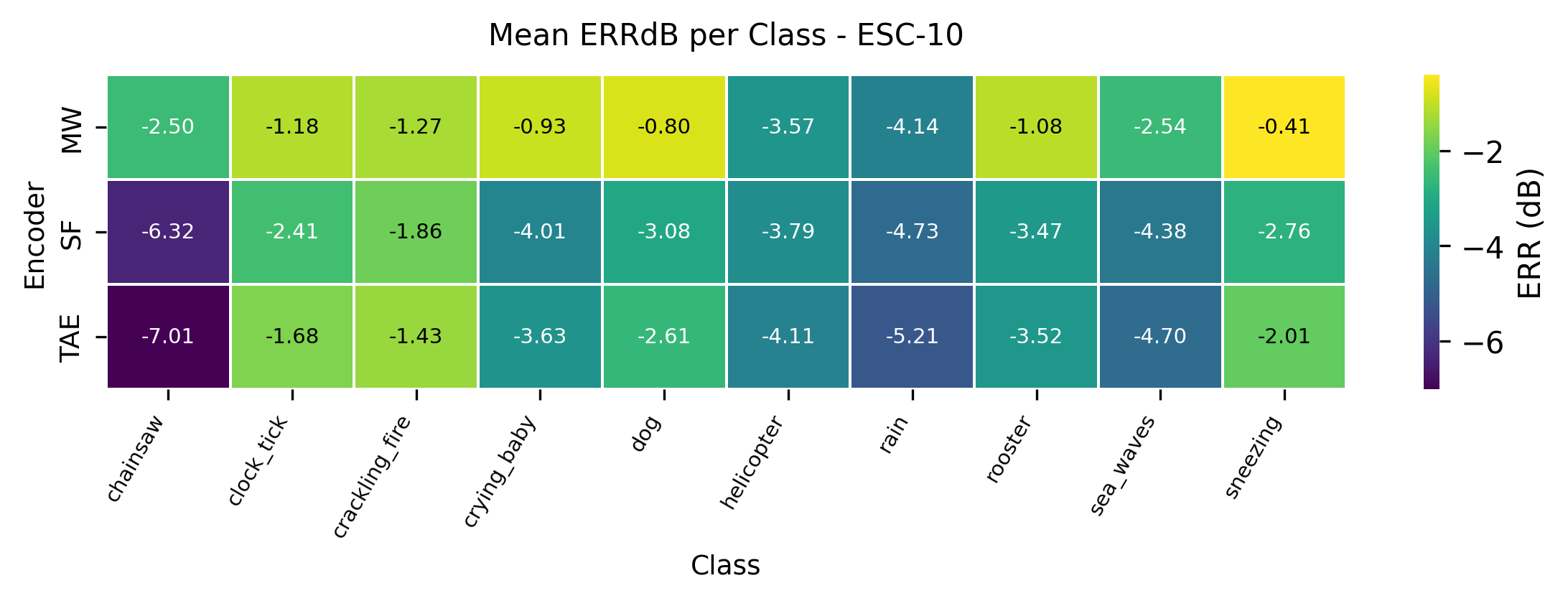}
\includegraphics[width=1.0\linewidth]{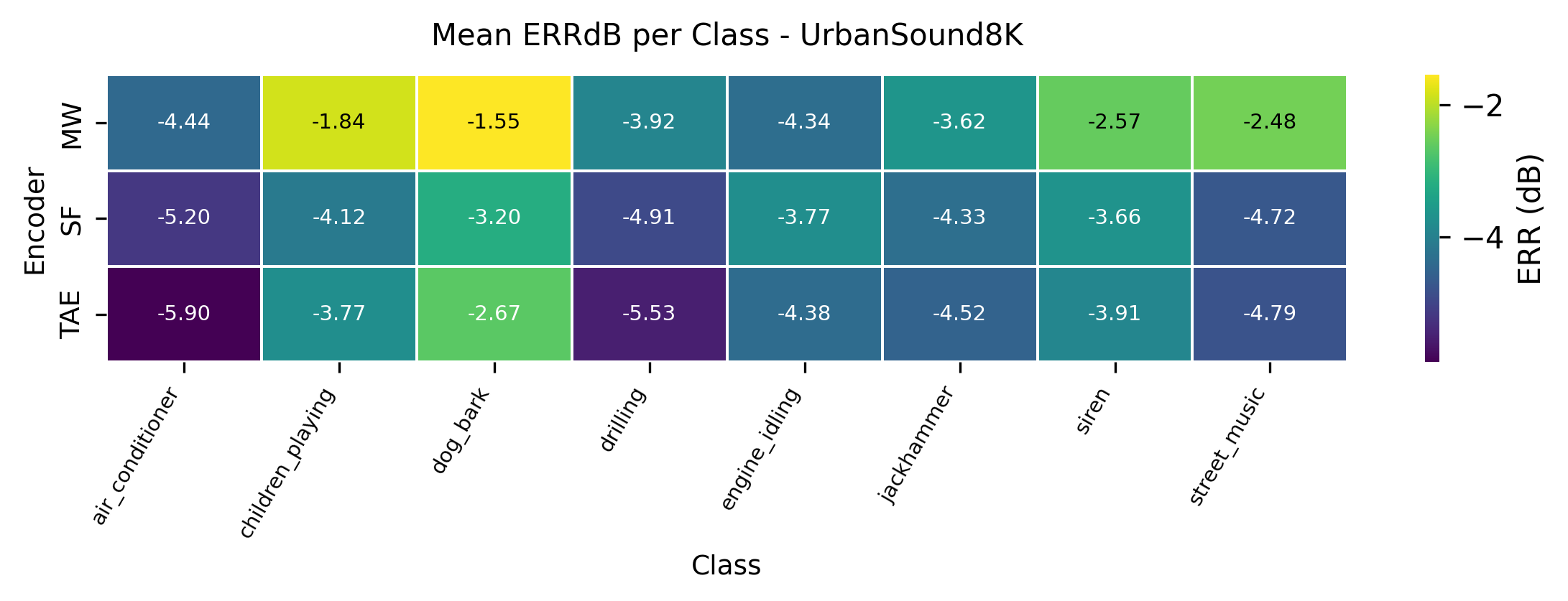}
\includegraphics[width=1.0\linewidth]{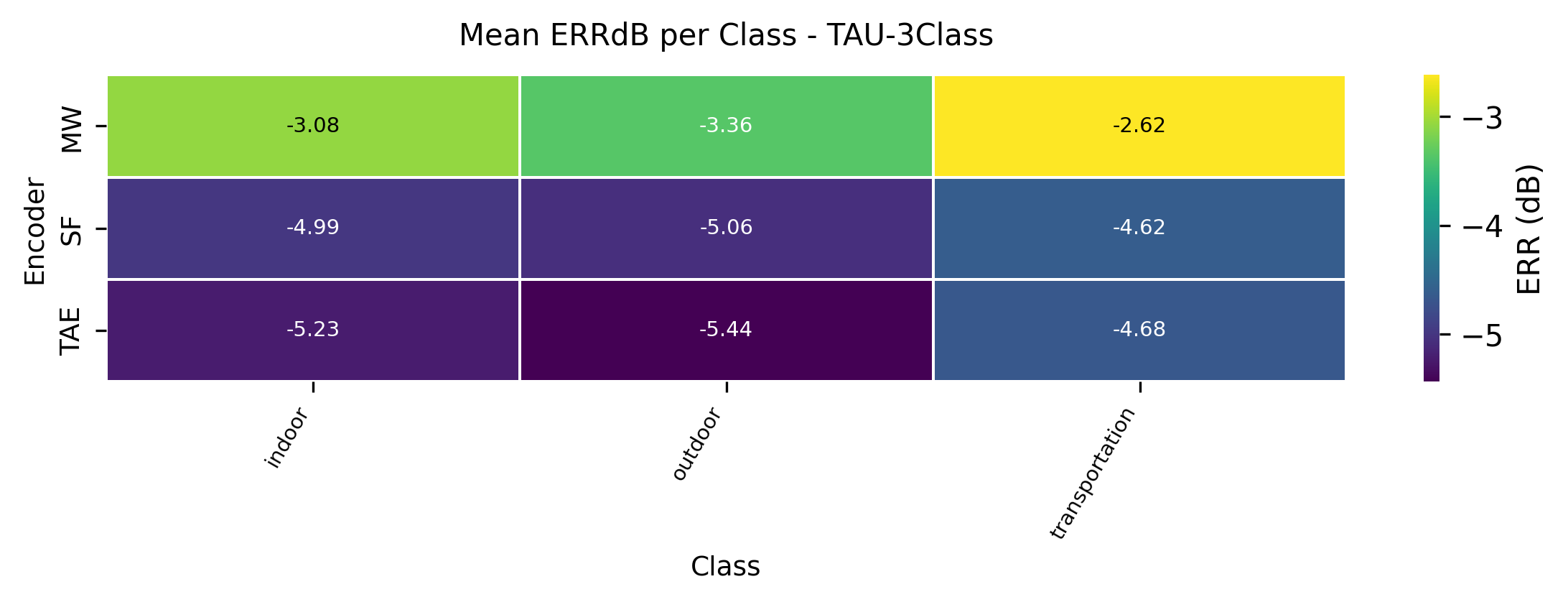}
\caption{Error in decibels (ERRdB) per class for each encoder across the evaluated datasets. MW: Moving Window, SF: Step Forward, TAE: Threshold Adaptive Encoder.}
\label{fig:ERRdB-categories}
\end{figure}

Figure~\ref{fig:ERRdB-categories} presents per-class ERRdB for each dataset. TAE consistently achieves the lowest reconstruction errors across most classes and scene types in all datasets, while MW tends to have the highest errors and SF remains intermediate. This pattern is observed for different class categories and encoders in the TAU-3Class, ESC-10, and UrbanSound8K datasets, indicating stable relative performance among the encoding methods.

TAE's adaptive thresholding yields reliable reconstruction of both stationary and temporally complex environmental sounds, while MW's windowing introduces the most artifacts. SF's compromise approach performs reasonably well but does not consistently match TAE, especially for scenes and complex classes.

\subsection{Encoding Efficiency}
\label{ssec:efficiency}

\begin{figure}[tb]
\centering
\includegraphics[width=0.9\linewidth]{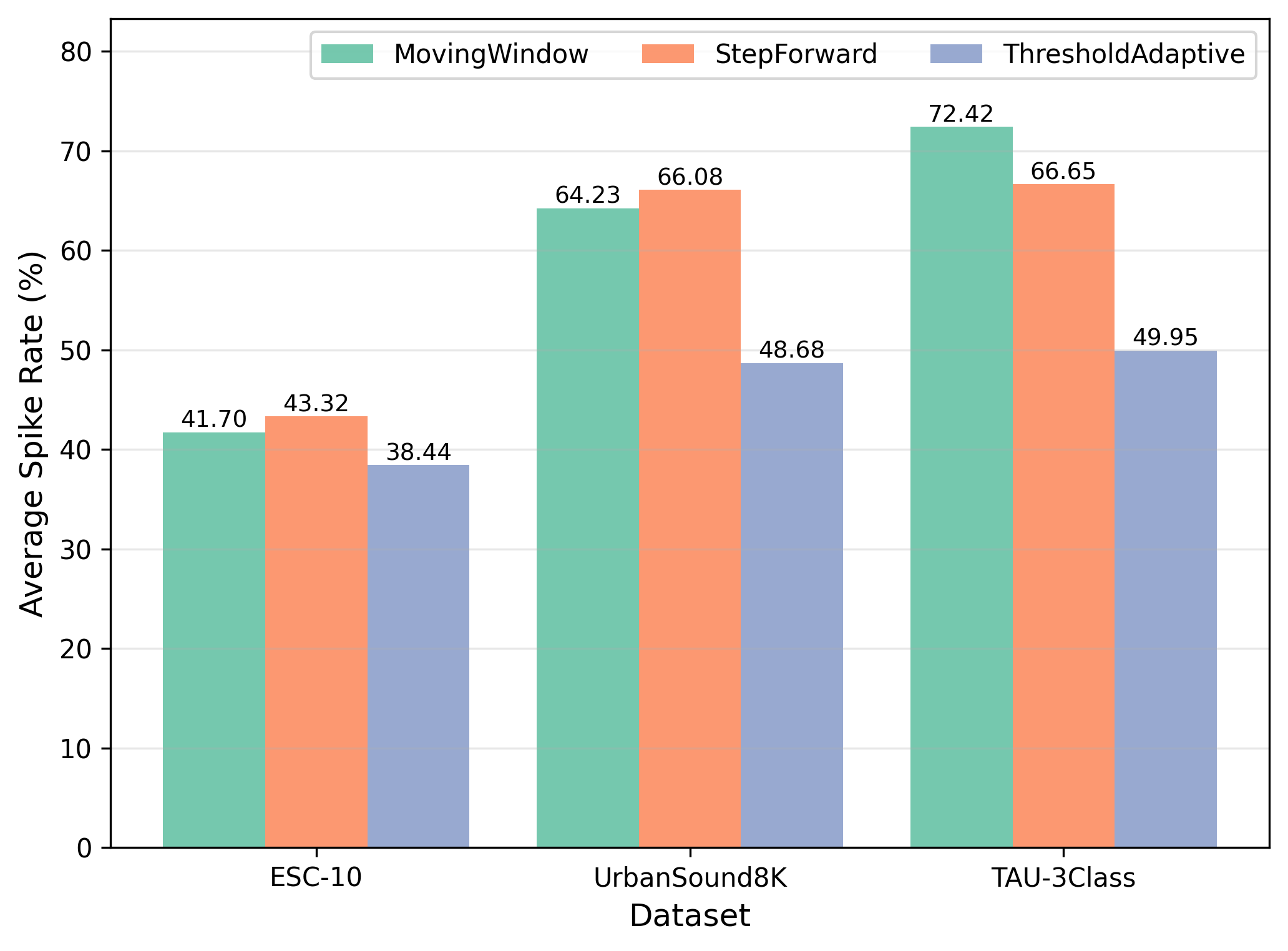}
\caption{Average spike firing rates across datasets and encoders. TAE yields the lowest spike firing rate in all cases, supporting efficient encoding.}
\label{fig:FiringRate-categories}
\end{figure}

Spike firing rate analysis (Figure~\ref{fig:FiringRate-categories}) indicates TAE consistently achieves the lowest average spike rates: 38.44\% (ESC-10), 49.95\% (TAU-3Class), and 48.68\% (UrbanSound8K), compared to MW (41.70\%, 72.42\%, 64.23\%) and SF (43.32\%, 66.65\%, 66.08\%). We also computed the average encoding time and memory usage for each encoder; both metrics are similar across methods (8.3–8.9 ms; ~0.16 MB), indicating that TAE’s advantage stems from reduced spike generation rather than differences in computational cost or memory footprint.

\subsection{Classification}
\label{ssec:classification}

Classification results of the trained SNN using each encoder method and dataset are summarized in Table~\ref{tab:classification-results}. TAE achieves the highest accuracy for ESC-10 and TAU-3Class (0.690), while SF is slightly better on UrbanSound8K (0.564). All spike encoders perform below the standard baselines provided by the original dataset references. However, it is important to note that the objective of the current study is to benchmark spike encoding methods rather than achieving the best classification performance, therefore the baseline values are only provided for reference. Future work should aim at developing specific SNNs suitable for environmental sound classification

\begin{table}[hbtp]
\caption{Encoder classification accuracy across datasets}
\centering
\small
\setlength{\tabcolsep}{4pt}
\begin{threeparttable}
\begin{tabular}{@{}lccc@{}}
\toprule
Encoder & ESC-10 & UrbanSound8K & TAU-3Class \\
\midrule
Moving Window & 0.620 & 0.530 & 0.550 \\
Step Forward & 0.598 & \textbf{0.564} & 0.640 \\
Threshold Adaptive & \textbf{0.690} & 0.535 & \textbf{0.690} \\
\midrule
Baseline & 0.727 & 0.730 & 0.873 \\
\bottomrule
\end{tabular}
\begin{tablenotes}[flushleft]
\footnotesize
\item Note: Baseline references values taken from: ESC-10~\cite{Piczak2015}; UrbanSound8K~\cite{salamon2014urbansound}; TAU-3Class (Task B)~\cite{heittola2020task1b}.
\end{tablenotes}
\end{threeparttable}
\label{tab:classification-results}
\end{table}

Our results highlight TAE as the most effective spike encoding method for both reconstruction and classification of environmental sounds, supporting its robust generalization and efficiency. This corroborates prior studies on adaptive thresholding for sequential data~\cite{Wang2024} and suggests future work should explore further architectural improvements for challenging sound recognition tasks.

\section{Discussion}
\label{sec:discussion}

This paper presents a comprehensive benchmark of spike encoding techniques specifically designed for environmental sound processing. Through extensive evaluation across three diverse datasets (ESC-10, UrbanSound8K, and TAU Urban Acoustic Scenes), we provide foundational insights that address a significant gap in neuromorphic audio processing research.

Our multi-faceted analysis reveals that Threshold Adaptive Encoding (TAE) consistently outperforms Moving Window (MW) and Step Forward (SF) methods across multiple evaluation criteria. TAE demonstrates superior signal reconstruction quality, achieving the lowest reconstruction errors across frequency bands and sound classes, while simultaneously maintaining the lowest spike firing rates, indicating enhanced energy efficiency. These findings establish TAE as the most suitable spike encoding approach for environmental sound applications where both accuracy and energy consumption are critical considerations. The superior performance of TAE aligns with previous findings on adaptive thresholding approaches in neuromorphic signal processing \cite{Wang2024, Petro2020}, extending their applicability from controlled environments to the more challenging domain of environmental acoustics.

The classification results further validate TAE's effectiveness, achieving the best performance on two of three evaluated datasets. While all spike encoders perform below traditional baselines, this performance gap is expected given the exploratory nature of applying SNNs to environmental sound classification. This gap highlights opportunities for future architectural improvements and corroborates findings from recent neuromorphic audio surveys \cite{Baek2024} that identify the need for specialized SNN architectures tailored to audio processing tasks. The performance gap is also consistent with observations in other domains where SNNs require careful co-design of encoding and network architecture to achieve competitive performance \cite{Eshraghian2023, Malcolm2023}.

Complementing this, SATRN (Spiking Audio Tagging Robust Network) shows that attention‑equipped SNN architectures can approach CNN‑like accuracy with efficiency on benchmarks including UrbanSound8K, indicating a clear path to close this gap \cite{SATRN2025}. Accordingly, a promising direction is to pair a encoder like TAE with attention‑based SNNs like SATRN to jointly optimize encoding fidelity and model capacity for noisy, overlapping scenes and resource‑constrained edge deployments \cite{SATRN2025,Baek2024}.

\vfill\pagebreak
\bibliographystyle{IEEEbib}
\bibliography{IEEEabrv}

\end{document}